\documentclass[aps,prb,twocolumn,showpacs,preprintnumbers,amsmath,amssymb,floatfix,superscriptaddress]{revtex4}
\usepackage{dcolumn}
\usepackage{bm}
\usepackage{amsmath}
\usepackage{feynmp}

\usepackage{graphicx}

\begin{document}

\title{Spin Conduction in Anisotropic $3-D$ Topological Insulators}

\author{Vincent E. Sacksteder IV}
 \email{vincent@sacksteder.com}
\affiliation{Institute of Physics, Chinese Academy of Sciences, Beijing 100190}

\author{Stefan Kettemann}
\affiliation{School of Engineering and Science, Jacobs University Bremen, Bremen 28759, Germany}
\affiliation{Division of Advanced Materials Science, Pohang University of Science and Technology (POSTECH), San 31, Hyoja-dong, Nam-gu, Pohang 790-784, South Korea}

\author{QuanSheng Wu}
\affiliation{Beijing National Laboratory for Condensed Matter
  Physics, and Institute of Physics, Chinese Academy of Sciences,
  Beijing 100190, China}

\author{Xi Dai}
\affiliation{Beijing National Laboratory for Condensed Matter
  Physics, and Institute of Physics, Chinese Academy of Sciences,
  Beijing 100190, China}

\author{Zhong Fang}
\affiliation{Beijing National Laboratory for Condensed Matter
  Physics, and Institute of Physics, Chinese Academy of Sciences,
  Beijing 100190, China}


\date{\today}

\begin{abstract}
When topological insulators possess rotational symmetry their spin lifetime is tied to the scattering time.   We show that  in anisotropic topological insulators this tie can be broken and the spin lifetime can be very large.  Two different mechanisms can obtain   spin conduction over long distances.  The first is tuning the Hamiltonian to conserve a spin operator $\cos \phi \, \sigma_x + \sin \phi \, \sigma_y$, while the second is tuning the Fermi energy to be near  a local extremum of the energy dispersion.     Both mechanisms can produce persistent spin helices.   We report spin lifetimes and spin diffusion equations.
\end{abstract}

\maketitle

\section{Introduction}

Topological insulators \cite{Fu07, Zhang09, Hasan10, Liu10b} exhibit a gap in the spectrum of bulk states, and bridging that gap is a band of surface states; if the Fermi energy is within the gap then electrons flow only  along the surface and not in the bulk. 
At small enough momenta the surface band has the shape of two cones joined at their ends. One Dirac cone describes electrons with positive energies, and the other describes negative energies.  An electron's spin is locked to its momentum, so backscattering is suppressed.    All these properties are consequences of time-reversal ($T$) symmetry, and are robust against small perturbations which are $T$-symmetric, such as non-magnetic impurities.

Recently much attention has been given to creating  topologically protected qubits at the interface between a 3-D topological insulator (TI) and a conventional superconductor, allowing robust quantum arithmetic \cite{Fu08}.  The TI spin-momentum locking  attracts attention to spintronics;   recent works have shown that circularly polarized light  could induce spin currents \cite{McIver11} and topological phase transitions \cite{Inoue10}.

This paper's main focus is on obtaining good spin conductors suitable for spintronics.  Disordered TIs are unusually poor spin conductors.    Because electronic spin is tied to  momentum,  each scattering event randomizes  the spin, as is  typical of Elliot-Yafet spin relaxation \cite{Wenk10}.  On the Dirac cone the spin lifetime $\tau_s$ is tightly coupled \cite{Burkov10} to the scattering time $\tau$ by $\tau_s / \tau = 2$.     Ordinary semiconductors have much longer spin lifetimes, because spin is conserved during scattering and is randomized only by precession between scattering events.  In this paper we will describe how to tune a TI  for very long spin lifetimes ($\tau_s \gg \tau$), allowing spin to conduct over long distances.  
 
  \begin{figure}[top]
\begin{center}
  \includegraphics[width=8cm, bb=0 0 400 200]{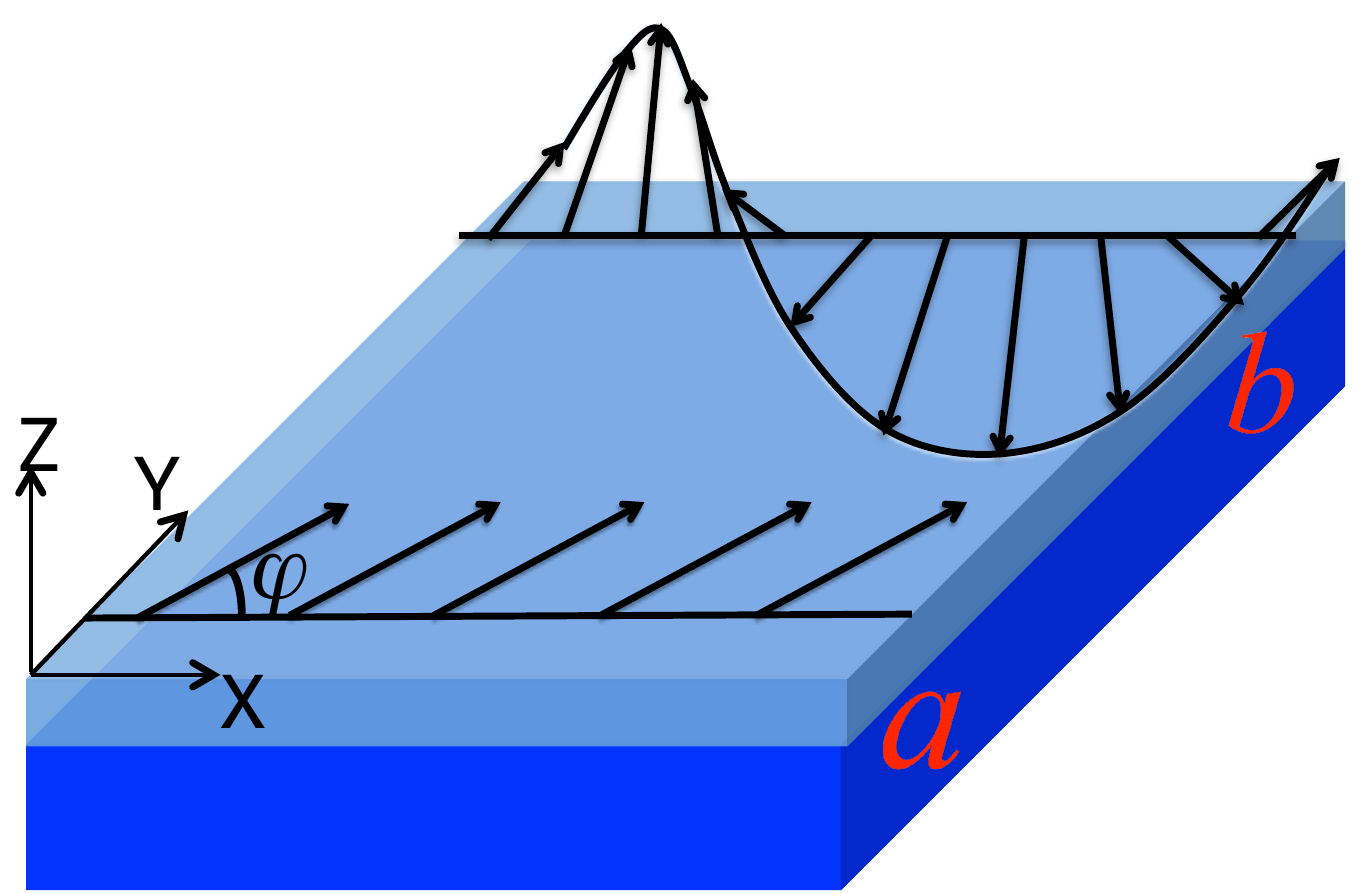} 
    \end{center}
    \caption{Two spin profiles that conduct  in properly tuned TIs.   The light blue and dark blue represent  the conducting surface and the  insulating bulk of a 3-D topological insulator.  Line  \textbf{\textit{a}} shows a spatially uniform spin density.    Line \textbf{\textit{b}} shows a persistent spin helix which oscillates along the x axis and is spatially uniform along the y axis.  The PSH is a standing wave; it does not precess.  }
    \label{fig:SpinTexture}
\end{figure}

We will show that the two spin profiles shown in  Figure   ~\ref{fig:SpinTexture}  conduct  in a properly tuned TI.   The first  profile lies in the surface plane and its angle $\phi$ in that plane is  constant.  The spin component $S_\phi$ aligned with $\phi$ conducts: its integral $\int {d\vec{x}} \,S_\phi, \, S_\phi \equiv \ cos \phi S_x + \sin \phi S_y$   is conserved  ($\tau_s = \infty$) and its  long-wavelength variations equilibrate diffusively, similarly to heat diffusion.  Spin at right angles to the unit vector $\hat{e}_\phi$ is filtered out very quickly, relaxing with lifetime $\tau_s = \tau$.  Figure   ~\ref{fig:SpinTexture}b shows the second spin profile which conducts in a properly tuned TI:  a standing spin wave  which repeats at intervals of $\pi / | \hat{Q} |$.   Because it rotates its spin orientation and does not decay, it is called a persistent spin helix (PSH) \cite{Bernevig06}.   Associated with the PSH is  a conserved quantity $\int {d\vec{x} }\,S_{PSH}, \; S_{PSH} = \cos( 2 \hat{Q} \cdot \vec{x}) (-\sin \phi S_x  + \cos \phi S_y) + \sin( 2 \hat{Q} \cdot \vec{x}) S_z$.  If $\phi = 0$ the PSH rotates in the $y-z$ plane, while if $\phi = \pi/2$ it rotates in the $x-z$ plane.  The spin orientation $\phi$, the wave-vector $  2 \hat{Q}$, and the diffusion constant all depend on the details of the TI Hamiltonian.

\begin{figure} [top]
\begin{center}
  \includegraphics[width=8cm, bb=0 0  500 250]{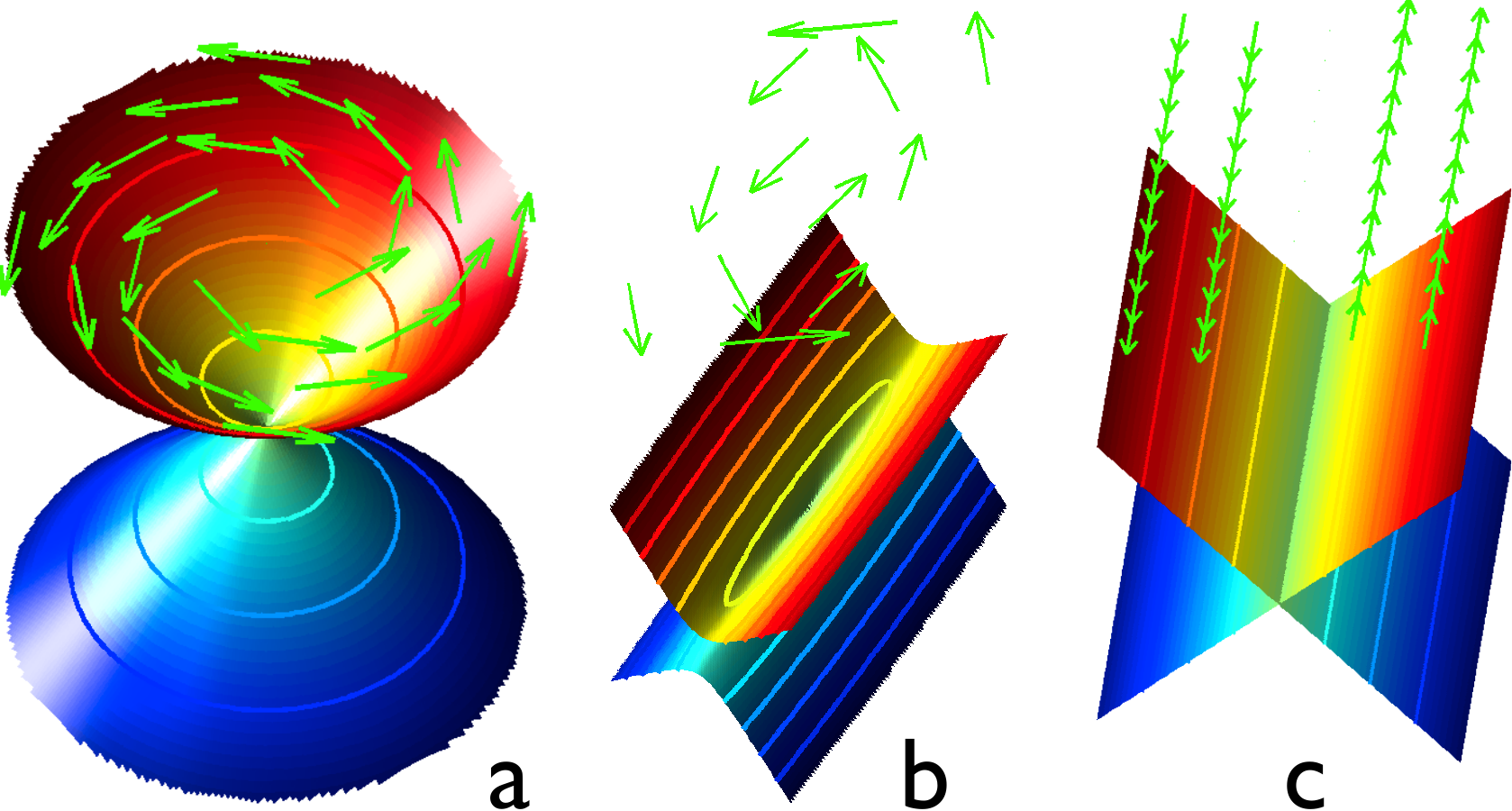}
\end{center}
   \caption{Anisotropy in TIs.  (a - isotropic)  the Rashba Hamiltonian $H_R=\hbar v_F(k_y\sigma_x - k_x \sigma_y)$, (b - stretched) linear terms from a $k \cdot p$ model of $\beta-Ag_2Te$, (c - infinitely stretched) a tuned model  $H_1 = \hbar v_F k_x (\cos \phi \, \sigma_x + \sin \phi \,\sigma_y)$ which  conserves the spin operator $\cos \phi \, \sigma_x + \sin \phi \, \sigma_y$.  The colored surfaces show the energy dispersion $E(\vec{k})$, and the lines inscribed on them show Fermi surfaces - contours of constant energy.  The arrows show the spin orientation, which is locked to the momentum.}
   \label{fig:Figure1}
\end{figure}

Tuning for spin conduction is not possible unless the surface band is more complex than a simple Dirac cone.   Anisotropy, i.e. violation of rotational invariance \footnote{Highly symmetric TI surfaces ($C_{n V}$, $n \geq 3$) require the Rashba Hamiltonian. Lower symmetries \cite{Oguchi09} (for instance $C_{2V}$) can produce anisotropic  Hamiltonians.}, is key to long distance spin conduction.  This is the origin of the spin orientation angle $\phi$ which parameterizes both  spin profiles.  Anisotropy can be realized in TIs either by choosing reduced-symmetry materials like $\beta - Ag_2 Te$  \cite{Zhang11, Virot11} (see Figure ~\ref{fig:Figure1}b) or by cutting a high-symmetry TI in a way that reduces the surface's symmetry \cite{Egger10,Moon11}.   Figure ~\ref{fig:Figure1} illustrates distortion of the Dirac cone as rotational symmetry is progressively broken.  We use simple linear models that are appropriate near the Dirac point.    Figure ~\ref{fig:Figure1}a shows the dispersion of a rotationally symmetric system. Figure ~\ref{fig:Figure1}b shows  the Dirac cone stretching along one axis as rotational symmetry is broken.  Finally Figure ~\ref{fig:Figure1}c extrapolates the stretching to its extreme: the energy dispersion depends only on $k_x$ not $k_y$, and therefore an operator $\cos \phi \, \sigma_x + \sin \phi \, \sigma_y$ is necessarily conserved.

\begin{figure}[top]
\begin{center}
  \includegraphics[width=8cm, bb=100 0 1100 550]{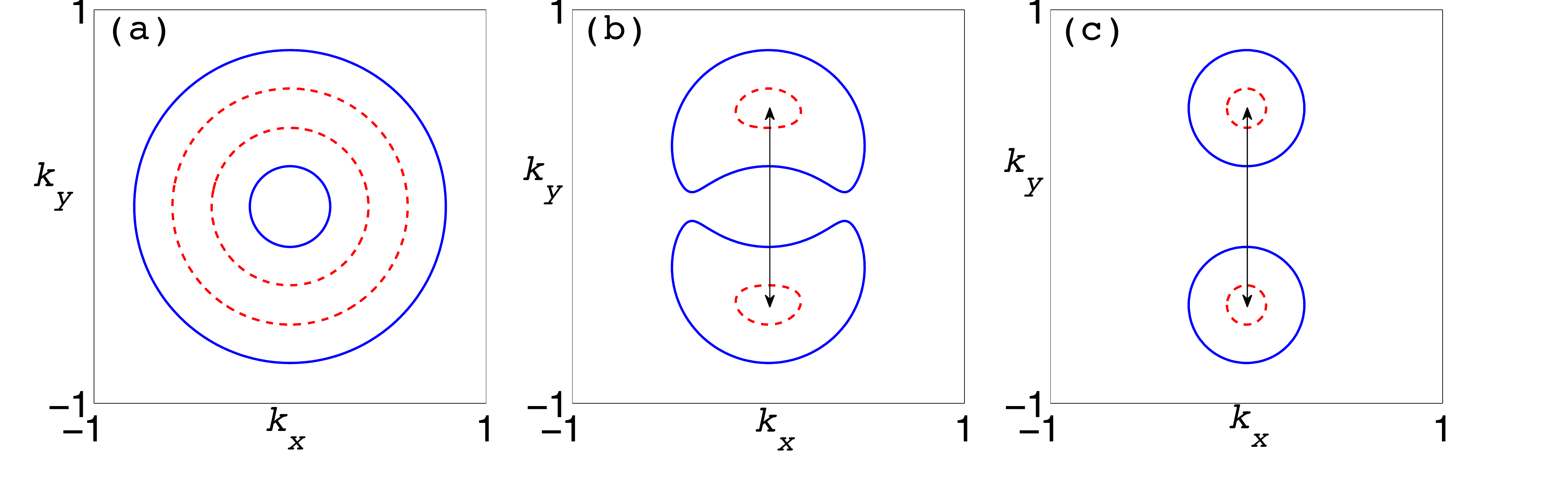} 
    \end{center}
    \caption{Fermi surfaces showing that anisotropy causes local extrema.   $H=\frac{k^2}{2m}+ v_x k_x\sigma_y+ v_y k_y\sigma_x, \, m=0.5,\, v_y=1$. The solid blue lines are Fermi surfaces at energy $E_F = -0.163$, while the dashed red lines are at $E_F =  - 0.24$. All quantities are unitless.  (a) Isotropic $v_x=1$, (b) anisotropic $v_x=0.8$, (c) spin conserving $v_x=0$.   (a) and (b) are typical of untuned spin-orbit couplings in quantum wells, while (c) is the same as Bernevig et al's model \cite{Bernevig06} of a tuned quantum well.  The dashed  Fermi surfaces in (b)  and (c) lie near local minima.   Model (b) conducts spin $S_x$ when $E_F \approx  E_Q$, while (c) always conducts $S_x$.  The arrows point out the Fermi surface nesting symmetry which causes PSH's in both models.  }
    \label{fig:Figure3}
\end{figure}

There are two ways to tune for spin conduction. The first is  to tune the Hamiltonian  for conservation of a spin operator $\cos \phi \sigma_x + \sin \phi \sigma_y$,   causing $S_\phi$  to  conduct.  This has been achieved  in GaAs quantum wells by tuning both the well width and the well depth to obtain partial cancellation of the Rashba and Dresselhaus terms \cite{Schliemann03,Bernevig06,Koralek09}.   These tuned quantum wells are modeled by a Hamiltonian that includes both a spin-conserving quadratic term $ \frac{\hbar^2 |\vec{k}|^2}{2m}$ and a small linear spin-orbit term which conserves $\cos \phi \sigma_x + \sin \phi \sigma_y$, as illustrated in Figure ~\ref{fig:Figure3}c.  They manifest a PSH-induced strong enhancement of the spin lifetime.  The  TI model of Figure ~\ref{fig:Figure1}c also conducts both $S_\phi$ and  PSH's with wave-vector $\pm 2 \hat{Q} = \pm 2 \frac{E_F}{v_F} \hat{x}$.  

We will show that there is a second way to obtain  spin conduction if the surface band has local extrema at $\vec{k} = \pm \hat{Q} \neq 0$, in which case tuning the  Fermi energy  near the energy $E_Q$ of the extrema will produce very long spin lifetimes.  Figure ~\ref{fig:Figure3} shows  that even a small anisotropy will produce the required local minima.  The model shown in Figure ~\ref{fig:Figure3}a includes a quadratic term and a spin-orbit term, both of which are rotationally symmetric.  It shows four Fermi surfaces -  contours of constant energy.  Two Fermi surfaces are at $E_F = -0.163$, two are at $E_F = -0.24$, and all four are perfect circles.  The minimum of the energy dispersion is also a circle located midway between the dotted $E_F = -0.24$ Fermi surfaces. In  Figure ~\ref{fig:Figure3}b the rotational symmetry is only slightly broken - the spin-orbit term's strength is only $20\%$ smaller along the $x$ axis than it is along the $y$ axis.   Nonetheless the  Fermi surfaces have already divided and wrapped themselves around the dispersion's local minima which are now two discrete points located inside the dotted $E_F = -0.24$ Fermi surfaces.   Model ~\ref{fig:Figure3}b does not conserve any spin operator, but we will show that it conducts spin when $E_F$ is adjusted so that the Fermi surfaces lie  close to the the local minima.     Lastly Figure ~\ref{fig:Figure3}c reduces the spin-orbit term along the $x$ axis to zero, and exhibits spin conduction at any value of $E_F$.     In  both Figures ~\ref{fig:Figure3}b and ~\ref{fig:Figure3}c spin conduction is associated with there being two disconnected  Fermi surfaces centered on the local extrema.  Spin conduction in TI's is possible only when anisotropy changes the global structure of the Fermi surface.

In order to understand this global physics, we calculate charge and spin conduction for a very general class of $T$-conserving spin-orbit Hamiltonians: $H_{gen} = a_x(\vec{k}) \sigma_x + a_y(\vec{k})\sigma_y + a_I(\vec{k})$, where $a_x, a_y$ are odd in $\vec{k}$ and $a_I$ is even.  This describes charge moving in the $x-y$ plane, on one surface of a 3-D TI .  We study the single-particle density matrix $\rho = \psi^\dagger \psi$,  which is a $2 \times 2$ matrix in spin space.  We write it as a $4-$vector $\vec{\rho} = \begin{bmatrix} N, & S_x, & S_y, & S_z \end{bmatrix} $ containing the charge density $N= {Tr}(\rho)$ and spin densities $S_x, S_y, S_z = {Tr}(\rho \sigma_i)/2$.   We will first  analyze $\vec{\rho}$'s structure, and later calculate its diffusion induced by disorder.  
   
   \section{Spin Profiles}
   
 The spatially uniform and PSH profiles shown in Figure ~\ref{fig:SpinTexture} manifest themselves  in the Fourier-transformed density matrix   $\rho(\vec{q}) \propto \int {d\vec{k}} \, \psi^{\dagger}(\vec{q}/2 + \vec{k}) \psi(-\vec{q}/2 + \vec{k})$ as strong peaks at $\vec{q} = 0$ and at $\vec{q} = \pm 2 \hat{Q}$.  These peaks  are directly linked to the TI Fermi surface: the state vector $\psi$ is populated only by states from the Fermi surface, and therefore $\rho(\vec{q})$ is peaked when $\vec{q}$ maximizes the intersection of the Fermi surface with a copy of itself shifted by $\vec{q}$.     (The dominance of the Fermi surface is assured if $E_F \tau / \hbar \gg 1$, the temperature is small, and there are no interactions.)  The spatially uniform peak $\rho(\vec{q} = 0)$ realizes this maximization trivially.  
 
Figures ~\ref{fig:Figure3}b and ~\ref{fig:Figure3}c illustrate a special nesting symmetry which produces PSH peaks at $ \vec{q} = \pm 2 \hat{Q}$.   They show pairs of Fermi surfaces  centered at $\pm \hat{Q}$ which possess inversion symmetry ($\hat{Q} \rightarrow - \hat{Q}$) because of the TI's $T$ symmetry.   The Fermi surfaces also possess nesting symmetry, which means that a shift of $\vec{q} = \pm  2 \hat{Q}$ moves one Fermi surface on top of the other.   This nesting symmetry produces peaks in $\rho(\vec{q})$ at $\vec{q} = \pm 2 \hat{Q}$; it is responsible for PSH's.   The nesting symmetry can be written as $\hat{Q} + \vec{k} \rightarrow -\hat{Q} + \vec{k}$, where $\hat{Q} + \vec{k}$ lies on one Fermi surface and $-\hat{Q} + \vec{k}$ lies on the other one.      $T$  plus nesting implies that the Fermi surface near $\hat{Q}$ possesses inversion symmetry around $\hat{Q}$: $\hat{Q} + \vec{k} \rightarrow \hat{Q} - \vec{k}$.

The  TI  surface has only one conduction band and only one valence band,  with  spin quantum number equal to   $s = +1, - 1$ respectively.  We assume that the Fermi surface lies only in the conduction band; $s = +1$.   As a result the spatially uniform $S_z$ spin density  is identically zero: $S_z(\vec{q} = 0)$'s contributing terms  are  of the form  $S_z \propto {Tr}(\rho_k \sigma_z )/2, \, \rho_k = \psi^{\dagger}(\vec{k}) \psi(\vec{k})$, which is zero for all $\psi$ in the conduction band.  Similarly the charge $N(\vec{q} = 2 \hat{Q})$ and spin $S_\phi(\vec{q} = 2 \hat{Q}) $ components   of the PSH are zero, because  $\rho(2 \hat{Q})  \propto \psi^{\dagger}(\vec{Q}) \psi(-\vec{Q})$ and  $T$ symmetry ensures that $N(2 \hat{Q}) = {Tr}(\rho(2 \hat{Q}))$  and $S_\phi = {Tr}(\rho(2 \hat{Q})\,(\cos \phi \, \sigma_x + \sin \phi \, \sigma_y)/2)$ are zero when $\psi$ is in the conduction band.  

\section{Diffusive Conduction}

 \begin{figure}[top]
\begin{center}
  \includegraphics[width=8cm, bb=0 0 480 220]{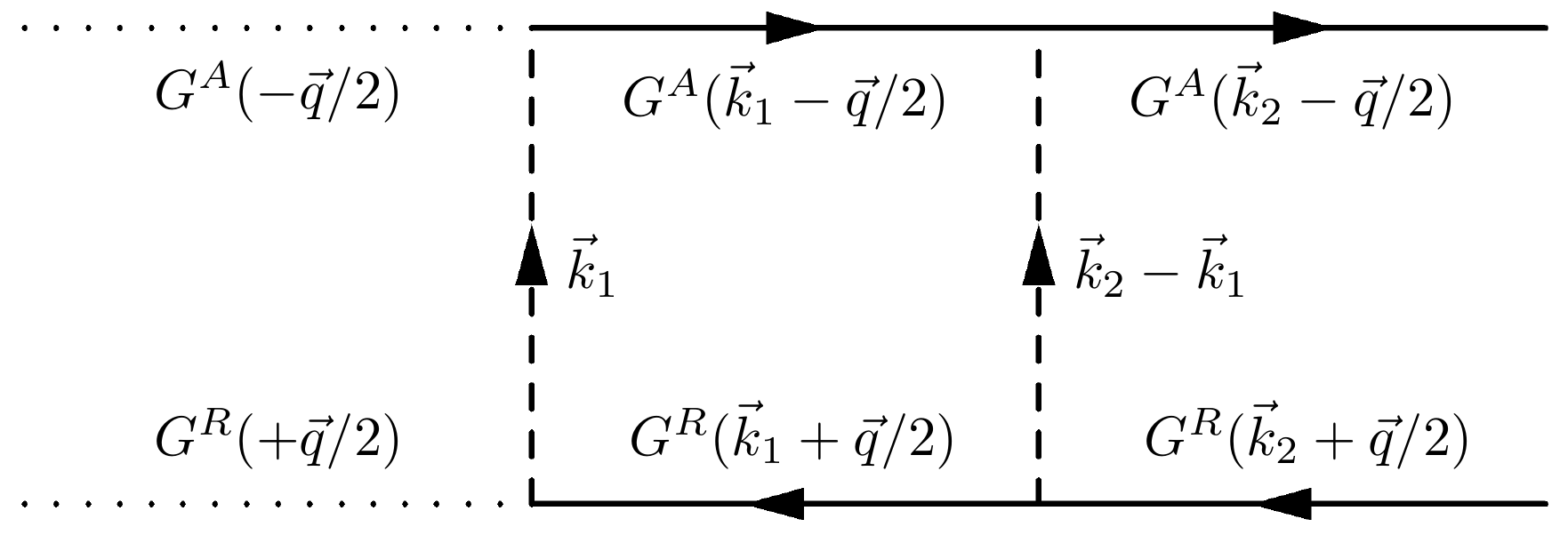} 
    \end{center}
    \caption{Diffusion diagram with two joint scatterings.  A single joint scattering event $I_{ij}$ corresponds to a pair of Green's functions $G^A$ and $G^R$ connected by  a dashed line. The dotted legs correspond to the external propagators.}
    \label{fig:JointScattering}
\end{figure}

We now consider adding a non-magnetic "white noise" disorder potential $V$ to the general Hamiltonian $H_{gen}$, where $V = \begin{bmatrix} 1 &  0 \\ 0 & 1 \end{bmatrix} u(\vec{r}), \, \langle u(\vec{r}) u(\acute{\vec{r}}) \rangle = n_i u_0^2 \delta(\vec{r} - \acute{\vec{r}})$, and $n_i u_0^2$ gives the disorder concentration and strength. When disorder is present  the density matrix evolves diffusively at time scales larger than the elastic scattering time.  Its evolution is controlled by the partial differential equation $\mathcal{D}_{ij}^{-1} \vec{\rho} = 0$,   where the  $4 \times 4$ matrix  $\mathcal{D}_{ij}$ is called the diffuson.  The diffuson's matrix structure couples the charge and spin densities to each other.   We  derive  the diffuson using standard methods from the diagrammatic technique for disordered systems \cite{Hikami80, Suzuura06, McCann06}, couched in the notation of References \onlinecite{Burkov04, Burkov10}.  Within this diagrammatic technique the conductivity is determined by the disorder-averaged two-particle correlation function, which is controlled by ladder diagrams at leading order in $(E_F \tau / \hbar)^{-1}$. The diffuson is composed of an infinite series of ladder diagrams like that seen in  Figure ~\ref{fig:JointScattering}.   These diagrams describe sequences of events in which $\psi$ and $\psi^\dagger$ move together, scattering in unison.    A single joint scattering event is described by the operator $I_{ij}$, and the diffuson  sums diagrams with any number of joint scatterings;  $ \mathcal{D}_{ij}(\vec{q}, \omega)  =  \sum_{n=0}^\infty (I_{ij})^n = (1 - I_{ij})^{-1}$. The joint scattering operator $I_{ij}$ is  pictured in Figure ~\ref{fig:JointScattering} and is given by the integral  
\begin{equation} 
\begin{split}
I_{ij} = \frac{n_i u_0^2}{2} \int d\vec{k} \, {Tr}(\, G^A( \vec{k} - \vec{q}/2, E_F)  \\ \sigma_i \, G^R(  \vec{k} + \vec{q}/2,  E_F + \hbar \omega) \, \sigma_j \, ) 
\end{split}
\label{JointScatteringOperator}
\end{equation}
$G^A$ and $G^R$  are the disorder-averaged single-particle Green's functions which express uncorrelated movements of $\psi$ and $\psi^\dagger$, while $\vec{q}$ is the diffuson momentum.  The trace is taken over the spin indices of $G^A, G^R, \sigma_i,$ and $\sigma_j$, which are all $2 \times 2$ matrices in spin space.

The zero-frequency  component of the diffuson $\mathcal{D}_{ij}( \vec{q} = 0, \omega = 0)$ is equal to $\hat{\tau}_s/ \tau$, where $\hat{\tau}_{s}$ is the  tensor  that  governs  relaxation of spatially uniform spin profiles.  Assuming as before that the Fermi surface is dominant ($E \tau / \hbar \gg 1$) and contains only the conduction band, we find:
\begin{equation}
\hat{\tau}_{s}/\tau =  2 \begin{bmatrix} 
0 & 0 & 0 
\\ 0 &(1 - \langle \cos 2 \theta \rangle_F )& - \langle \sin 2 \theta \rangle_F 
\\ 0 & - \langle \sin 2 \theta \rangle_F & (1 + \langle \cos 2 \theta \rangle_F)
\end{bmatrix}^{-1}
\end{equation}
This result is fully general for all $H_{gen}$.  The angle $\theta(\vec{k})$ gives the relative strength of the $a_x \sigma_x$ and $a_y \sigma_y$ terms, and is defined by $\tan \theta =  a_y/ a_x$.     The average $\langle \; \rangle_F$ is over the entire Fermi surface(s), and is weighted by the density of states.     The zeros mean that the charge lifetime is infinite; charge is conserved.  Linear combinations  of $S_x, S_y$ have lifetimes $2 \tau / (1 \pm \sqrt{\langle \cos 2 \theta \rangle_F^2 + \langle \sin 2 \theta\rangle_F^2} )  $;  spin conduction is obtained only  if
\begin{equation} 1 = {\langle \cos 2 \theta \rangle_F^2 + \langle \sin 2 \theta \rangle_F^2}\,. \label{Masslessq0Condition}
\end{equation}
 
\begin{figure}[top]
\begin{center}
  \includegraphics[width=8.5cm, height=4cm, bb= 0 0 300 150]{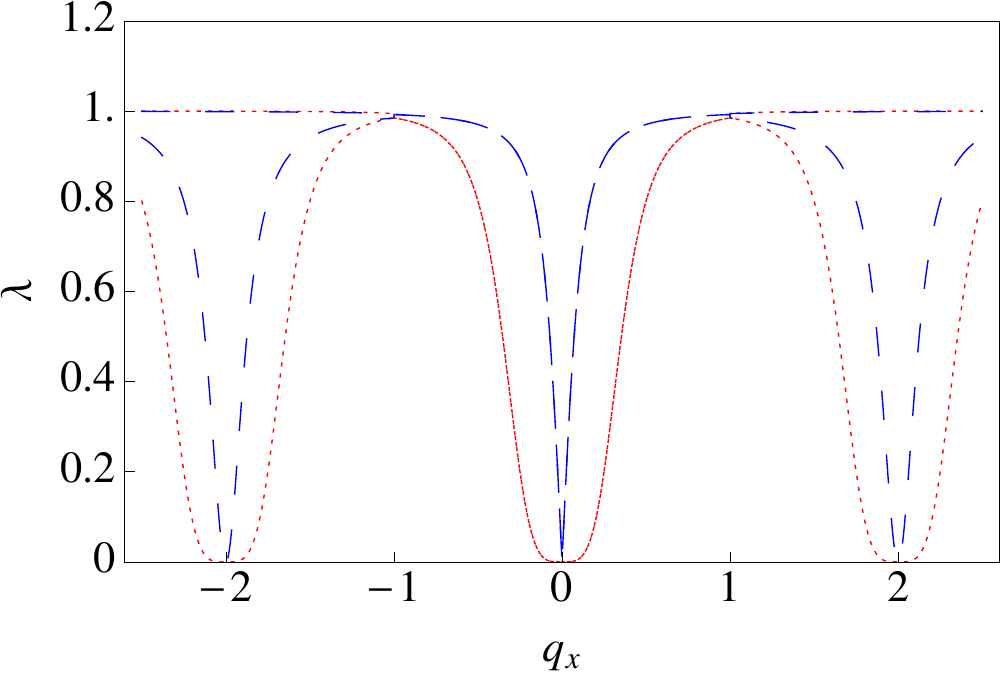}
   \end{center}
   \caption{Smallest two eigenvalues (absolute values) of the inverse diffuson  as a function of momentum. Zeros occur both at $\vec{q}=0$ and at $\vec{q} = \pm 2 \hat{Q}$, implying conduction of both the spatially uniform spin profile and PSH's. Dashes and dots are for the linear and quadratic models seen in Figures ~\ref{fig:Figure1}c and ~\ref{fig:Figure3}c respectively.   The Fermi energy $E_F$ has been tuned to match the local minima seen in Figure ~\ref{fig:Figure3}c.  The momentum scale is  $\hat{Q} = |\hat{Q}| \hat{x}$, and $\tau \, E_F = 8$. }
   \label{FigurePSHb}
\end{figure}

 Equation \ref{Masslessq0Condition} confirms our earlier statement that spin conduction can be obtained by tuning for conservation of a spin operator $\cos \phi \, \sigma_x + \sin \phi \, \sigma_y$.  In this case  $| \cos \theta |, \, | \sin \theta |$ are equal to the constants $| \cos \phi |, \, | \sin \phi |$ and  equation \ref{Masslessq0Condition} is trivially satisfied.      Figure ~\ref{FigurePSHb} shows the spectrum of the inverse diffuson in two models which conserve $\cos \phi \, \sigma_x + \sin \phi \, \sigma_y$.   Both models have two null eigenvalues at $q_x = 0$, corresponding to conduction of charge and of $S_\phi$.  Both models also exhibit a single null eigenvalue at $q_x = \pm 2 \hat{Q}$, implying PSH conduction.   Models which both  conserve  $\cos \phi \, \sigma_x + \sin \phi \, \sigma_y$ and have a nesting symmetry with $\theta(\hat{Q}) = \phi$  always exhibit PSH's.  When the nesting symmetry is only approximate, the PSH lifetime is $ {\tau_{PSH}}  = \hbar^2 / (4 \tau \langle  (\delta E)^2 \rangle_F), \, \delta E = \vec{\xi} \cdot \nabla_k E(\hat{Q})$.   The gradient in this formula measures violation of the inversion symmetry $\hat{Q} + \vec{k} \rightarrow \hat{Q} - \vec{k}$.  $\vec{\xi} =  \vec{k} - \hat{Q}$ is the displacement of the Fermi surface from the extremum.  
  
Equation \ref{Masslessq0Condition} also confirms that tuning the Fermi energy can produce spin conduction.   The key is that this equation concerns only the Fermi surface: $|\cos \theta |$ and $ |\sin \theta |$ must be constant there.   When  $E_F$ is tuned close to a local extremum the Fermi surface becomes very small. Therefore $|\cos \theta |, \, |\sin \theta |$ are nearly constant on the Fermi surface, $\cos \phi \, \sigma_x + \sin \phi \, \sigma_y, \, \phi = \theta(\hat{Q})$ is nearly conserved there, and  $S_\phi$ conducts freely.  

Tuning the Fermi energy also can produce  PSH's. $T$ symmetry requires that  extrema always come in pairs at $\vec{k} = \pm \hat{Q}$.  The Fermi surfaces $\mathcal{S}_\pm$ accompanying these pairs possess the nesting symmetry which produces PSH's.  However there must be no scattering between the pair of Fermi surfaces $\mathcal{S}_\pm$ and any other Fermi surfaces, because $\cos \phi \, \sigma_x + \sin \phi\, \sigma_y, \, \phi = \theta(\hat{Q})$ will not be conserved on the other surfaces.  If there are additional Fermi surfaces then the disorder potential must be smooth, without short-wavelength variations.  In this case  there will be one pair of persistent spin helices for each pair of extrema.    

 We have computed the spin  lifetime  when there are local extrema.      For the spatially uniform spin profile it is $\tau_s/ \tau =   1 / \langle (\delta \theta)^2 \rangle_F, \, \delta \theta = \vec{\xi} \cdot \nabla_k \theta(\hat{Q})$.  The PSH lifetime is double this value.    Our PSH calculation is valid only in the diffusive regime where  the PSH characteristic length $ l_{PSH} =  h / 2 | \hat{Q} |$ is large compared to the scattering length $l_\tau$. $\delta \theta$ measures the amount that $\theta$ varies on the Fermi surface, because when $\theta$ is constant $S_\phi$ is conserved and $\tau_s = \infty$.     $\vec{\xi}$ measures the width of the Fermi surface;  when $E_F$ approaches the extremum it  goes to zero and the spin lifetime diverges.    For instance, in the  quadratic model $H=\frac{k^2}{2m}+v_x k_x\sigma_y+ v_y k_y\sigma_x, \, v_y > v_x$ shown in Figure ~\ref{fig:Figure3}  the lifetime is   $\tau_s = \tau \frac{2 m v_y^2  (v_y^2 - v_x^2)}{v_x^2 (E_F - E_Q) } $.   It diverges when spin is conserved ($v_x = 0$)  and also when the Fermi energy $E_F$ is tuned to the extremum $E_Q$.   When the model is tuned for rotational symmetry $v_x \rightarrow v_y$  the lifetime becomes very small because the local minimum becomes very shallow, the Fermi surface stretches along the $x$ axis, and $\langle \xi_x^2 \rangle_F \propto (E_F- E_Q) / (v_y^2 - v_x^2)$ becomes very large.

   Local extrema have already been realized in a TI \cite{Hsieh08, Zhang09a}:  $Bi_{1-x}Sb_{x}$, which has six fold symmetry.  ARPES measurements  \cite{Hsieh08} reveal at least three six-fold degenerate minima, with momenta at  $|\hat{Q}| \approx 0.15, 0.8, 1.1 \text{\AA}^{-1}$.    However the symmetry is too high:  substitutional disorder causes scattering between all six minima.  Moreover the bulk gap is very small, and the PSH length scale is so short that it may lie in the ballistic regime.   
   
  Local extrema will be found  whenever there is an avoided band crossing in an anisotropic material.   In $Bi_{1-x}Sb_{x}$ a conventional surface band occurs very close to the TI band.  Repulsion between these two bands causes the observed local minima.  Avoided band crossings have also been observed in very thin TI films - the TI bands on each of the film's two surfaces couple to each other, causing band repulsion and extrema  \cite{Lu10, Linder09, Liu10}.   The remaining necessary ingredient for spin conduction is anisotropy.   In this respect the recent predictions of 10 to 1 anisotropy \cite{Zhang11} in $\beta - Ag_2 Te$ and 18 to 1 anisotropy in metacinnabar \cite{Virot11} are very encouraging.
        
When spin conducts  - for instance when  $E_F$ is tuned near an extremum -  the magnetoresistance  will become null or even change sign.  If only the spatially uniform profile conducts then there will be neither weak localization nor antilocalization (null magnetoresistance).  If  there are PSH's then there will be  a complete reversal from weak antilocalization to weak localization, from positive to negative magnetoresistance.

Returning to equation \ref{JointScatteringOperator}, we have  calculated the diffuson operator $\mathcal{D}_{ij}$ which controls spin diffusion via $\mathcal{D}_{ij}^{-1}\vec{S} = 0$.  Our calculation is general for all $H_{gen}$ but considers only long wavelengths; i.e. momenta near $\vec{q} = 0$.  For brevity we will present here only the result when  $\cos \phi \, \sigma_x + \sin \phi \, \sigma_y$ is conserved. The spin component orthogonal to $\hat{e}_\phi$ decays with lifetime $\tau_s = \tau$, and we have already seen that $S_z = 0$. The spin diffusion equation for $N$ and $S_\phi$ is:
\begin{eqnarray}
 &\,&  \begin{bmatrix}\partial_t - \frac{1}{2}\nabla_x \cdot D \cdot \nabla_x & -2 \nabla_x \cdot \vec{\Gamma}   \\ -\frac{1}{2} \nabla_x \cdot \vec{\Gamma}  & \partial_t - \nabla_x \cdot D \cdot \nabla_x  \end{bmatrix} \begin{bmatrix} N \\ S_\phi \end{bmatrix} = 0
 \nonumber \\
& \vec{\Gamma}  & = \frac{1}{\hbar } \langle \nabla_k E(\vec{k}) \rangle_F, \; D = \frac{\tau}{\hbar^2 } \langle \nabla_k E(\vec{k}) \otimes \nabla_k E(\vec{k}) \rangle_F
\label{SpinConservingUniforma}
\end{eqnarray}
 The charge-spin coupling $\nabla_x \cdot \vec{\Gamma}$ and the diffusion tensor $D$  are determined entirely by the energy dispersion.  If the Fermi surface is an ellipse with height $h$ and width $w$ then $\vec{\Gamma} = 0$ and $D = 8 \tau (E_F - E_Q)^2/ \hbar^2 \, (w^{-2}\hat{x}\otimes \hat{x} + h^{-2} \hat{y} \otimes \hat{y})$. 
Assuming that the PSH length scale $ l_{PSH} =  h / 2 | \hat{Q} | \gg l_\tau$ is in the diffusive regime, we have derived  the PSH diffusion equation:
     \begin {equation} 
  ( \partial_t    + 1/\tau_{PSH}  
-     (\nabla_x - 2 \imath \hat{Q}) \cdot D \cdot (\nabla_x - 2 \imath \hat{Q})) S_{PSH} = 0
 \label{SpinConservingPSHa}
\end{equation}
 The term with two $ (\nabla_x - 2 \imath \hat{Q}) $ derivatives implies that small deviations from the spin helix  relax diffusively.     

\section{Conclusion}
In this article we studied spin conduction in a very general model of TI surfaces with non-magnetic disorder.  We calculated the spin decay times  and spin diffusion equations and found two ways to tune for a long spin lifetime and spin conduction.   The first tuning mechanism is well known from quantum wells but new to TIs: tuning the Hamiltonian to conserve a spin operator.  We found a second tuning mechanism: tuning the Fermi energy near a local extremum of the energy dispersion.  Neither mechanism is possible unless the TI surface is anisotropic.  Both mechanisms cause conduction of a spatially uniform spin profile. If the Fermi surfaces exhibit an additional  nesting symmetry then Persistent Spin Helices will also conduct.   When spin conduction is realized the TI's magnetoresistance will be either null or negative, unlike an untuned TI where the magnetoresistance is positive. TIs which combine anisotropy with avoided band crossings will be promising candidates for spin conduction and PSH's.

 We acknowledge support from the NSF of China (Grant No. NSFC 10876042 and
No. NSFC 10874158),  the 973 program of China (Grant No. 2007CB925000 and No. 2011CBA00108)), and the WCU (World Class University) program of POSTECH through R31-2008-000-10059-0, Division of Advanced Materials Science.  V. E. S. acknowledges the hospitality of AMS, POSTECH.

\bibliography{Vincent}
\end{document}